\documentstyle{laa}
\begin{document}
\thesaurus{08.01.1 08.16.3 10.01.1 10.08.1}

\title{Abundances of four very metal-poor stars of the BPS survey}
\author{Francesca Primas$^1$,  Paolo Molaro$^2$, Fiorella Castelli$^3$}
\offprints{P. Molaro, Osservatorio Astronomico di Trieste, Via G.B.
Tiepolo 11, Trieste, Italy}
\institute{$^1$ Dipartimento di Astronomia, Universit\`a
degli Studi di Trieste, I--34131, Trieste, Italy \\
$^2$ Osservatorio Astronomico di Trieste, Via G.B. Tiepolo 11,
I--34131, Trieste, Italy \\
$^3$ C.N.R.-G.N.A. and Osservatorio Astronomico di Trieste,
Via G.B. Tiepolo 11, I--34131, Trieste, Italy}

\date{Received date; accepted date}
\maketitle

\begin{abstract}

High resolution CASPEC and EMMI spectra (R$\approx$10$^4$) of
the faint metal-deficient candidates CS 22891--209, CS 22897--8,
CS 22948--66 and CS 22968--14 taken from the survey of Beers,
Preston and Shectman (1985, BPS) are analyzed for
abundances. The fine analysis with LTE model atmospheres gives
an iron metallicity for the four stars between $-$3.15 and
$-$3.45 in good agreement with the abundances inferred from
CaII K index by Beers et al. (1992). These four objects
considerably augment the number of spectroscopically studied
stars at the low metallicity extreme.

We found the $\alpha$-elements Mg, Ca, Si, and Ti enhanced,
and the iron-peak elements Sc, Mn and Cr tracking closely iron
with the typical pattern of the more metallic halo stars.
Aluminum results deficient with respect both to iron and to
magnesium showing a strong {\it odd-even\/} effect, but for
[Fe/H] $\leq$ $-$3.0 [Al/Mg] levels out into a plateau around
$-$1.0, showing a primary behaviour. [Na/Fe] is $\approx$ $-$0.2
and only compared to magnesium it shows a mild odd-even effect.
The heavy elements Sr and Ba show a large scatter which is
not observed for other elements and cannot be due to
observational errors only. In the case of CS 22891--209 they
are particularly overabundant with [Sr,Ba/Fe] $\approx$
0.9. The spread might reflect the presence of strong
inhomogeneities in the interstellar gas at the earliest stages
of Galaxy formation. This might be an important signature of
the still elusive processes of the early formation of these
elements observed in old stars. \\

\end{abstract}

\section{Introduction}

Important aspects of the history of the Galaxy are recorded in
the chemical evolution which is controlled by the formation rate
and lifetimes of the stars responsible for the element formation.
The abundances we observe today result from the cumulative effects
of a number of stellar generations, and the interpretation is
rather complex. The less processed material contained in the
atmospheres of the oldest stars of the Galaxy provides a simpler
context to probe nucleosynthesis theories and models of chemical
evolution. Just a few supernovae events might be in fact
responsible for the elements observed in the most metal-poor stars.
A 25\,M$_{\rm \sun}$ supernova with an iron ejecta of
0.2--0.3\,M$_{\rm \sun}$ is able to pollute the 10$^6$\,M$_{\rm \sun}$
required for the gravitational instability to a metallicity of
[Fe/H] $\approx$ $-$4.8, which is shortly below the lowest value
observed in the more metal-deficient stars. Ryan et al. (1991)
discovered few metal-poor stars virtually free of heavy elements
suggesting that they belong to the {\it second} stellar generation
formed out by gas that experienced only one metallic production. \\
A flat metallicity distribution function at the lowest measurable
abundances gives a very low statistical weight to stars virtually
free of metals, which are therefore very rare (Beers, 1987).
Recent years registered a significant increase in the discoveries
of very metal-poor objects mainly achieved thanks to the HK
objective-prism/interference-filter survey of Beers et al. (1985).
The follow-up high-resolution spectroscopy of the most metal-poor
candidates confirmed that they are indeed very metal-poor objects:
CS 22876--32 with [Fe/H] = $-$4.29 (Molaro and Castelli, 1990),
CS 22885--96 and CS 22881--39 with [Fe/H] = $-$4.21 and $-$3.53
respectively (Molaro and Bonifacio, 1990), CS 22884--108 with
[Fe/H] = $-$3.27 (Norris et al., 1993). These stars complemented
three objects found by other techniques: CD$-$38$^{\circ}$245 with
[Fe/H] = $-$4.50 (Bessel and Norris, 1984), G64--12 with
[Fe/H] = $-$3.50 (Carney and Peterson, 1981) and G275--4 with
[Fe/H] = $-$3.70 (Ryan et al., 1991). The metallicities of these
objects are at the lowest levels ever observed in the universe,
about two orders of magnitude less than the oldest globular clusters
and lower than the metallicities of Damped and Metal-line systems at
the highest observable redshifts. Here, we report new chemical abundance
determinations for other four BPS objects  selected among the most
metal-poor candidates. For all of them we confirm an extremely low
metallicity at a level of [Fe/H] $\leq$ $-$3.15, thus further increasing
the number of spectroscopically determined ultra metal-deficient stars.
A preliminary report was already given in Molaro et al. (1993).

\begin{table}
\caption{Journal of observations}
\begin{flushleft}
\begin{tabular}{lrccc} \hline
     &                          &       &               &     \\
Star & \multicolumn{1}{c}{Date} & Range & T$_{\rm exp}$ & S/N \\
     &                          & \AA   & sec           &     \\
     &                          &       &               &     \\  \hline
                     &         &            &      &    \\
CD$-$38$^{\circ}245$ &  1/9/90 & 3900--5300 & 4320 & 40 \\
CD$-$38$^{\circ}245$ & 21/9/91 & 4500--6800 & 3600 & 80 \\
CS 22891--209        &  1/9/90 & 3900--5300 & 2700 & 50 \\
CS 22897--8          &  1/9/90 & 3900--5300 & 5408 & 50 \\
CS 22948--66         &  1/9/90 & 3900--5300 & 7200 & 50 \\
CS 22968--14         &  1/9/90 & 3900--5300 & 6714 & 40 \\
CS 22968--14         & 21/9/91 & 4500--6800 & 6000 & 40 \\
                     &         &            &      &    \\ \hline
\end{tabular}
\end{flushleft}
\end{table}

\section{Observations and data reduction}

The observations have been collected during two observing runs
on September 1990 and September 1991 and details are given in
the journal of Table 1.
The stars have been taken from the BPS survey on the ground of
their particularly low metallicity. In Table 2 the relevant data
for the stars, identified by their plate number, are summarized.
The photometric indices are taken from Beers et al. (1992) with
exception for CS 22968--14, for which the more recent data from
Norris et al. (1993) have been assumed. Furthermore, we derived
E(U$-$B) from E(B$-$V) using the reddening ratio E(U$-$B)/E(B$-$V)
= 0.72 + 0.05E(B$-$V) (Lang, 1992).
The first run on September 1990 was with the 3.6m telescope of
the European Southern Observatory (ESO), La Silla, equipped with
CASPEC. The 31.6 lines/mm Echelle grating was used in combination
with the 300 lines/mm cross-disperser. The CCD was a Tektronix
512$\times$512 pixels, 27 microns wide, thinned and backside
illuminated (ESO n$\fdg$16). The spectra are centered at 4550\AA~
covering the spectral region between 3900\AA~ and 5300\AA.
In the second run on September 1991 we used EMMI spectrograph of
the 3.5m NTT ESO telescope. The Echelle grating 10 with grism 5
as a cross disperser have been used. The detector was the CCD ESO
n$\fdg$24, FORD 2048$\times$2048 pixels, 15 microns wide. The
spectra cover the spectral range from 4500\AA~ to 6800\AA.

The data have been reduced using the Echelle package of Midas,
which includes order searching, flat fielding, background
subtraction, cleaning of cosmics and wavelength calibration.
Th-Ar lamp images were taken before and after each exposure.
The unidimensional spectra calibrated in wavelengths have been
normalized through a spline of the continuum points for each
order. The two configurations provide a resolution of about
11.5 and 15\,km\,s$^{-1}$ for EMMI and CASPEC spectra
respectively. Heliocentric radial velocities have been derived
from the CASPEC calibrated spectra with an accuracy of about
$\pm$1.5\,km\,s$^{-1}$. For CS 22891--209, CS 22897--8,
CS 22948--66 and CS 22968--14 we have derived 87, 262, $-$168.5,
and 162\,km\,s$^{-1}$ respectively.
\scriptsize
\begin{table}
\caption{The studied stars and photometric data}
\begin{flushleft}
\begin{tabular}{lcccccc}\hline
 & & & & & & \\
STAR               &    RA      &     DEC     &   V   & B-V  & U-B  & E(B-V)\\
 & & & & & & \\ \hline
                   &            &             &       &      &      &       \\
$-$38$^{\circ}$245 & 00 44 12   & $-$37 56 04 & 12.00 & 0.81 & ...  & 0.01  \\
891--209           & 19 37 38.6 & $-$61 10 49 & 12.17 & 0.83 & 0.24 & 0.05  \\
897--8             & 20 58 55.9 & $-$65 17 07 & 13.33 & 0.69 & 0.10 & 0.00  \\
948--66            & 21 41 48.4 & $-$37 41 45 & 13.47 & 0.63 & 0.01 & 0.00  \\
968--14            & 03 05 03.6 & $-$54 42 03 & 13.72 & 0.74 & 0.08 & 0.015 \\
 & & & & & & \\ \hline
\end{tabular}
\end{flushleft}
\end{table}
\normalsize

\subsection{Equivalent Widths}

Equivalent widths have been measured for all the lines that
resulted unblended by means of a synthetic spectrum tecnique.
They are listed in Table 3\footnote{Table 3 is available in electronic
form via an anonymous ftp copy at the CDS.}. The uncertainty in the
measure of the EW is estimated with  $W=3\sigma$FWHM. This is also
the minimum detectable equivalent width in case of non detection.
Our S/N ratios are tipically between 20 and 60, and the minimum
detectable feature ranges from 10 to 40 m\AA.

The procedure has been tested using our observations of
CD$-$38$^{\circ}$245, which has been extensively studied.
This star has been observed  in both the observing sessions
covering a spectral region of about 2900\AA, from 3900\AA~
to 6800\AA. Figure 1 shows the comparison between our EWs
with those measured by Bessel and Norris (1984), Molaro and
Castelli, (1990), Gratton and Sneden (1988), and Peterson and
Carney (1989). The comparison does not reveal systematic
effects, with a possible small offset of the measures of
Peterson and Carney with all the others.
\begin{figure}
\picplace{8.5cm}
\caption{Comparison between our determinations of EWs and those of
Bessel and Norris (1984, {\it triangles\/}), Gratton and Sneden (1988,
{\it spiked squares\/}), Peterson and Carney (1989, {\it squares\/})
and Molaro and Castelli (1990, {\it pentagons\/})}
\end{figure}

\begin{table*}
\caption{table3}
\vspace{22cm}
\end{table*}

\begin{table}
\caption{Parameters of our sample}
\begin{flushleft}
\begin{tabular}{lcccccc} \hline
 & & & & & & \\
Star&Teff$^{1}$&Teff$^{2}$&$\logg^{1,2}$&$\xi$&Teff$^{*}$&$\logg$^{3}$ \\
 & K & K & cm\,s$^{-2}$ & km\,s$^{-1}$ & K & cm\,s$^{-2}$\\
 & & & & & & \\ \hline
                &      &      &      &     &      &      \\
891--209        & 4890 & 4900 & 1.20 & 2.0 & 4800 & 1.40 \\
897--8          & 5060 & 5050 & 1.50 & 1.9 & 4925 & 1.65 \\
948--66         & 5178 & 5170 & 1.80 & 1.9 & 4950 & 1.20 \\
968--14         & 4945 & 4950 & 2.00 & 1.9 & 4920 & 1.25 \\
                &      &      &      &     &      &      \\ \hline
\end{tabular}
\end{flushleft}
$^{\rm (1)}$~~ Values deduced from the UBV photometry; \ $^{\rm (2)}$~~
Adopted values; \ $^{\rm (3)}$~~ Values deduced from the ionization balance;
$^{\rm (*)}$~~ Values deduced from the excitation temperatures. \\
\end{table}

\section{Spectroscopic analysis}

Abundance analysis has been performed following usual methods,
which are based on the comparison of observed and computed
equivalent widths and line profiles. More details can be found
in Castelli and Hack (1988).
No special effort has been done to derive the model parameters
for CD$-$38$^{\circ}$245 for which we take T$_{\rm eff}$ = 4700\,K
and $\log g$ = 1.4 from Gratton \& Sneden (1988).
For the other stars, the model parameters T$_{\rm eff}$ and $\log g$
were obtained by fitting the observed U$-$B and B$-$V colours,
reported in Table 2 and corrected for reddening, to the computed
colours given in Appendix\,A.
According to the metallicities estimated by Beers et al. (1992),
we used the [M/H]=$-$3.0 grid for CS 22891--209, CS 22897--8,
CS 22948--66, and the [M/H]=$-$3.5 grid for CS 22968--14. In these
grids the abundances of the $\alpha$-elements are enhanced by
0.4\,dex. Convection was treated with the mixing-length theory,
taking the mixing-length to the pressure scale-height ratio l/h
equal to 1.25. No overshooting was taken into account.
We computed for each star, in a consistent way with the models
described in more details in Appendix\,A, ATLAS9 models having the
adopted parameters given in Table 4.
Microturbulent velocities and abundances were derived by using the
WIDTH code (Kurucz, 1993b), which computes equivalent widths by
using line and continuum opacities integrated through the input
model atmosphere. For each element a guessed starting abundance is
varied little by little to get the best agreement between the calculated
and the measured equivalent widths. \\
The value of the microturbulent velocity $\xi$ was set to minimize any
dependence of iron abundance from the equivalent widths of Fe\,{\sc i}
lines. Only lines which give abundances differing by less than 1$\sigma$
from the mean were considered.
They can be found in Table 3. The total number of lines for each star
is given in brackets in the fourth column of Table 5.
By varying the microturbulent velocity at steps of 0.1\,km\,s$^{-1}$
we found $\xi$ = 1.9\,km\,s$^{-1}$ for all the stars of Table 2, except
for CS 22891--209 for which $\xi=2.0$\,km\,s$^{-1}$ was determined.
For CD$-$38$^{\circ}$245 the microturbulent velocity has been found of
2.0\,km\,s$^{-1}$, the same value of Molaro and Castelli (1990) and
Gratton and Sneden (1988) but lower than the 2.5\,km\,s$^{-1}$ by
Peterson et al. (1990) and the 3.5\,km\,s$^{-1}$ by Bessel and Norris
(1984). \\
We checked the adopted temperatures and gravity for each star. By
using the numerous Fe\,{\sc i} lines we derived the model temperature
structure which gives the minimum deviation of the abundance from the
excitation potential of the lower level. We found effective
temperatures which are systematically lower by about 150--200\,K than the
color-based ones, with the only exception of CS 22968--14, for which
the two determinations give approximately the same value. These
effective temperatures (T$_{\rm eff}^{*}$) are reported in the sixth column
of Table 4. \\
We checked the surface gravities by imposing the agreement between the
abundances derived from Fe\,{\sc ii} and Fe\,{\sc i} lines. The grid of
models used for each star is given in Table 5. The gravities which give
the best match between Fe\,{\sc i} and Fe\,{\sc ii} abundances agree within
$\pm$0.15-0.75\,dex with those derived from colours and are given in
the last column of Table 4.
Both effective temperatures derived from excitation temperatures and
gravities derived from ionization equilibria are sensitive to
microturbulent velocity and to errors in the equivalent widths.
Another uncertainty comes from the possibility of overionization, as
has been pointed out by Nissen et al. (1993) in the halo star HD 140283.
Neglecting NLTE effects should give a lower abundance from Fe\,{\sc ii}
lines and from low-excitation Fe\,{\sc i} lines. All these uncertainties
could explain the different model parameters obtained by using different
approaches. \\
We remark that for CS 22968--14 we derived T$_{\rm eff}$ equal to 4950\,K,
which is 150\,K higher than that obtained by Norris et al. (1993), the
same value $\log g$ = 2.0 for the gravity, and a microturbulent velocity
$\xi$=1.9\,km\,s$^{-1}$, lower by 0.6\,km\,s$^{-1}$ than that of Norris
et al. (1993).
\scriptsize
\begin{table}
\caption{Matches performed for gravity check}
\begin{flushleft}
\begin{tabular}{lccccr} \hline
 & & & & & \\
Star&$Teff&$\logg$&$\logA_{FeI}$&$\logA_{FeII}$&\multicolumn{1}{c}{$\DeltaA}\\
 & & & & & \\ hline
 & & & & & \\
          &      & 1.00 & $-$7.62$\pm$0.15 (51) & $-$7.78$\pm$0.18 (9) & 0.16\\
891--209  & 4900 & 1.20 & $-$7.64$\pm$0.15 (51) & $-$7.72$\pm$0.18 (9) & 0.08\\
          &      & 1.50 & $-$7.67$\pm$0.16 (51) & $-$7.63$\pm$0.18 (9) &-0.04\\
 & & & & & \\
          &      & 1.25 & $-$7.77$\pm$0.15 (46) & $-$7.93$\pm$0.23 (6) & 0.15\\
897--8    & 5050 & 1.50 & $-$7.79$\pm$0.15 (46) & $-$7.85$\pm$0.23 (6) & 0.06\\
          &      & 1.75 & $-$7.81$\pm$0.15 (46) & $-$7.77$\pm$0.23 (6) &-0.04\\
 & & & & & \\
          &      & 1.00 & $-$7.58$\pm$0.21 (52) & $-$7.64$\pm$0.20 (6) & 0.06\\
948--66   & 5170 & 1.25 & $-$7.60$\pm$0.19 (52) & $-$7.57$\pm$0.20 (6) &-0.03\\
          &      & 1.50 & $-$7.62$\pm$0.19 (52) & $-$7.49$\pm$0.20 (6) &-0.13\\
          &      & 1.80 & $-$7.65$\pm$0.19 (52) & $-$7.40$\pm$0.20 (6) &-0.25\\
 & & & & & \\
          &      & 1.00 & $-$7.85$\pm$0.17 (58) & $-$7.94$\pm$0.22 (9) & 0.09\\
968--14   & 4950 & 1.50 & $-$7.90$\pm$0.17 (58) & $-$7.78$\pm$0.22 (9) &-0.12\\
          &      & 2.00 & $-$7.94$\pm$0.16 (58) & $-$7.62$\pm$0.22 (9) &-0.32\\
 & & & & & \\ \hline
\end{tabular}
\end{flushleft}
\end{table}
\normalsize
\small
\begin{table}
\caption{Sensitivity of Sr, Ba and Fe abundances to stellar parameters
and EWs}
\begin{flushleft}
\begin{tabular}{lccccc} \hline
 & & & & & \\
 & \multicolumn{2}{c}{[Sr/H]} & \multicolumn{2}{c}{[Ba/H]} & [Fe/H] \\
 & & & & & \\
 &   4077\AA   &   4215\AA    &    4554\AA   &    4934\AA  &        \\
 & & & & & \\ \hline
                                    &       &       &       &       &       \\
{\bf CD$-$38$^{\circ}$245\/}        &       &       &       &       &       \\
$\Delta T_{\rm eff}$ (+200\,K)      &  0.16 &  0.16 &  0.18 &  0.18 &  0.26 \\
$\Delta T_{\rm eff}$ ($-$200\,K)    & -0.15 & -0.17 & -0.19 & -0.19 & -0.28 \\
$\Delta \log g$ (+0.25\,dex)        &  0.05 &  0.06 &  0.07 &  0.07 &  0.04 \\
$\Delta \log g$ ($-$0.25\,dex)      & -0.04 & -0.06 & -0.06 & -0.06 & -0.04 \\
$\Delta \xi$ (+0.5\,km\,s$^{-1}$)   & -0.33 & -0.13 & -0.02 & -0.01 & -0.18 \\
$\Delta \xi$ ($-$0.5\,km\,s$^{-1}$) &  0.50 &  0.25 &  0.03 &  0.02 &  0.25 \\
$\Delta$EW (+3$\sigma$)             &  0.35 &  0.28 &  0.46 &  0.64 &  0.15 \\
$\Delta$EW ($-$3$\sigma$)           & -0.33 & -0.24 &  ...  &  ...  & -0.15 \\
                                    &       &       &       &       &       \\
{\bf CS 22891--209\/}               &       &       &       &       &       \\
$\Delta T_{\rm eff}$ (+200\,K)      &  0.21 &  0.20 &  0.21 &  0.20 &  0.23 \\
$\Delta T_{\rm eff}$ ($-$200\,K)    & -0.19 & -0.20 & -0.22 & -0.22 & -0.24 \\
$\Delta \log g$ (+0.25\,dex)        &  0.03 &  0.03 &  0.03 &  0.05 & -0.03 \\
$\Delta \log g$ ($-$0.25\,dex)      & -0.01 & -0.01 & -0.04 & -0.06 &  0.02 \\
$\Delta \xi$ (+0.5\,km\,s$^{-1}$)   & -0.22 & -0.20 & -0.47 & -0.45 & -0.16 \\
$\Delta \xi$ ($-$0.5\,km\,s$^{-1}$) &  0.16 &  0.12 &  0.60 &  0.51 &  0.24 \\
$\Delta$EW (+3$\sigma$)             &  0.33 &  0.21 &  0.35 &  0.31 &  0.15 \\
$\Delta$EW ($-$3$\sigma$)           & -0.51 & -0.30 & -0.46 & -0.36 & -0.15 \\
                                    &       &       &       &       &       \\
{\bf CS 22897--8\/}                 &       &       &       &       &       \\
$\Delta T_{\rm eff}$ (+200\,K)      &  0.13 &  0.16 &  0.15 &  0.19 &  0.18 \\
$\Delta T_{\rm eff}$ ($-$200\,K)    & -0.19 & -0.17 & -0.19 & -0.18 & -0.28 \\
$\Delta \log g$ (+0.25\,dex)        &  0.00 &  0.06 &  0.05 &  0.08 & -0.06 \\
$\Delta \log g$ ($-$0.25\,dex)      & -0.08 & -0.05 & -0.08 & -0.06 & -0.02 \\
$\Delta \xi$ (+0.5\,km\,s$^{-1}$)   & -0.46 & -0.43 & -0.19 & -0.22 & -0.21 \\
$\Delta \xi$ ($-$0.5\,km\,s$^{-1}$) &  0.48 &  0.48 &  0.37 &  0.44 &  0.19 \\
$\Delta$EW (+3$\sigma$)             &  0.76 &  0.65 &  0.33 &  0.37 &  0.15 \\
$\Delta$EW ($-$3$\sigma$)           & -0.95 & -0.82 & -0.30 & -0.31 & -0.15 \\
                                    &       &       &       &       &       \\
{\bf CS 22948--66\/}                &       &       &       &       &       \\
$\Delta T_{\rm eff}$ (+200\,K)      &  0.11 &  0.17 &  0.16 &  0.17 &  0.18 \\
$\Delta T_{\rm eff}$ ($-$200\,K)    & -0.16 & -0.15 & -0.17 & -0.17 & -0.29 \\
$\Delta \log g$ (+0.25\,dex)        &  0.05 &  0.08 &  0.08 &  0.08 & -0.06 \\
$\Delta \log g$ ($-$0.25\,dex)      & -0.06 & -0.06 & -0.07 & -0.07 & -0.02 \\
$\Delta \xi$ (+0.5\,km\,s$^{-1}$)   & -0.39 & -0.34 & -0.05 & -0.02 & -0.24 \\
$\Delta \xi$ ($-$0.5\,km\,s$^{-1}$) &  0.50 &  0.50 &  0.08 &  0.04 &  0.20 \\
$\Delta$EW (+3$\sigma$)             &  0.60 &  0.64 &  0.38 &  0.30 &  0.19 \\
$\Delta$EW ($-$3$\sigma$)           & -0.62 & -0.59 & -0.47 & -0.45 & -0.19 \\
                                    &       &       &       &       &       \\
{\bf CS 22968--14\/}                &       &       &       &       &       \\
$\Delta T_{\rm eff}$ (+200\,K)      &  0.15 &  0.16 &  0.18 &  0.18 &  0.20 \\
$\Delta T_{\rm eff}$ ($-$200\,K)    & -0.17 & -0.16 & -0.17 & -0.18 & -0.29 \\
$\Delta \log g$ (+0.25\,dex)        &  0.07 &  0.08 &  0.08 &  0.07 & -0.06 \\
$\Delta \log g$ ($-$0.25\,dex)      & -0.08 & -0.07 & -0.07 & -0.07 & -0.02 \\
$\Delta \xi$ (+0.5\,km\,s$^{-1}$)   & -0.07 & -0.04 & -0.01 & -0.01 & -0.18 \\
$\Delta \xi$ ($-$0.5\,km\,s$^{-1}$) &  0.11 &  0.07 &  0.02 &  0.01 &  0.16 \\
$\Delta$EW (+3$\sigma$)             &  0.73 &  0.72 &  0.58 &  0.63 &  0.16 \\
$\Delta$EW ($-$3$\sigma$)           & -0.90 &  ...  &  ...  &  ...  & -0.16 \\
 & & & & & \\ \hline
\end{tabular}
\end{flushleft}
\end{table}
\normalsize
\section{Abundances}

Abundances derived from the  equivalent widths of lines
measured in our spectra are reported in Table 3, together with
the average value and the dispersion around the mean. For
completeness, in the last column of Table 3, the solar abundances
from Anders and Grevesse (1989) and for iron from Hannaford et al.
(1992) are reported as a fraction of the total number of atoms N(t)
($\log$ N(H)/N(t) = $-$0.04).
The $\log gf$ values reported in Table 3 are taken from the line
lists provided by Kurucz (1993c), supplemented by the critically
compiled iron-peak {\it gf}-values of Fuhr, Martin and Wiese (1988)
and Martin, Fuhr and Wiese (1988). The unweighted mean of the
dispersion gives the size of random errors associated with the
measure of individual lines together with uncertainties in the
{\it gf}--values. For Fe\,{\sc i}  the dispersion around the mean
abundance amounts to 0.25--0.30\,dex (1$\sigma$), which is on the
same order of the dispersion found by Norris et al. (1993) for giant
stars with comparable signal-to-noise data. In addition systematic
errors may come from the choice of atmospheric parameters. A
detailed analysis of the sensitivity of abundances to errors in
atmospheric parameters is given in Norris et al. (1993). An
estimate of the dependence of the iron abundance from the
atmospheric parameters is given in Table 6 for all the stars.
The variation is sensitive to effective temperature, ionization
stage and gravity, being different for different elements as
detailed in Norris et al. (1993). In most cases the variation is
in the same direction for the various elements cancelling each
other in relative abundances. Major difficulties concern elements
for which the number of identifications is very small as silicon
and calcium. Obviously, larger errors are possible in these cases.
An explicit calculation has been performed for the elements Sr and
Ba, as reported in Table 6. The abundances have been later confirmed
by means of synthetic spectra generated with the SYNTHE code (Kurucz,
1993c). The total broadening has been assumed of 11\,km\,s$^{-1}$ for
EMMI or 14\,km\,s$^{-1}$ for CASPEC spectra. The consistency between
the required broadening and the instrumental broadening implies that
the rotational velocities are of few km\,s$^{-1}$ at most. Some
examples of synthetic spectra are shown in Fig.\,2 and in Fig.\,3
where we plotted the magnesium {\it b\/} multiplet for the stars
CS 22897--8 and CS 22968--14 and the sodium lines region for
CS 22968--14 and CD$-$38$^{\circ}$245 respectively.\\
\begin{figure}
\picplace{9.5cm}
\caption{Magnesium b multiplet region of CS 22897-8 and CS 22968-14:
the synthetic spectrum (continuous line) is superimposed to the data
(histogram)}
\end{figure}
\begin{figure}
\picplace{9.5cm}
\caption{Sodium lines region of CS 22968-14 and CD $-$38$^{\circ}$245}
\end{figure}

\section{Discussion}

Table 7 gives the general picture of all the relative abundances
we determined for our sample of stars. For CS 22891--209,
CS 22897--8, CS 22948--66 and CS 22968--14 the derived
metallicities are [Fe/H] = $-$3.15$\pm$0.15, $-$3.30$\pm$0.15,
$-$3.16$\pm$0.19 and $-$3.45$\pm$0.16. The agreement is good
with Beers et al. (1992) who estimated $-$2.93, $-$3.08, $-$3.11
and $-$3.52 respectively.
In Figs.\,4,6,7 and 8 we reported the relative abundances of our stars
compared with those of the literature available for halo stars
([Fe/H]$\leq$ $-$1.0). For the stars with [Fe/H] $\leq$ $-$3.0,
CD $-$38$^{\circ}$245, CS 22968--14, G64--12, CS 22876--32 and
CS 22885--96, we connected the abundance determinations of various
authors. Special care has been used to eliminate
offsets originated by different solar abundances among various
authors. Hereafter all the abundances given in Fig.\,4,6,7 and 8
are normalized to the solar scale of Anders and Grevesse (1989)
and Hannaford et al. (1992) for iron.
\begin{table}
\caption{Relative elemental abundances}
\begin{flushleft}
\begin{tabular}{lrrrrr} \hline
 & & & & & \\
 & $-$38$^{\circ}$245 & 891--209 & 897--8 & 948--66 & 968--14 \\
 & & & & & \\ \hline
          &         &         &         &         &          \\
$[Fe/H]$  & $-$4.13 & $-$3.15 & $-$3.30 & $-$3.16 & $-$3.45  \\
          &         &         &         &         &          \\
$[Na/Fe]$ & $-$0.16 &  ...    &  ...    &  ...    & $-$0.20  \\
          &         &         &         &         &          \\
$[Mg/Fe]$ & 0.41    & 0.73    & 0.23    & 0.01    & 0.02     \\
          &         &         &         &         &          \\
$[Al/Fe]$ & $-$0.80 & $-$0.03 & $-$0.68 & $-$1.26 & $-$0.87  \\
          &         &         &         &         &          \\
$[Si/Fe]$ & 0.39    & 0.45    & 0.36    &  ...    &   ...    \\
          &         &         &         &         &          \\
$[Ca/Fe]$ & 0.29    & 0.43    & $-$0.12 & $-$0.10 & 0.01     \\
          &         &         &         &         &          \\
$[Sc/Fe]$ & $-$0.03 & 0.01    & 0.04    & 0.16    & 0.12     \\
          &         &         &         &         &          \\
$[Ti/Fe]$ & 0.24    & 0.42    & 0.11    & 0.25    & 0.59     \\
          &         &         &         &         &          \\
$[Cr/Fe]$ & $-$0.63 & $-$0.16 & $-$0.09 & $-$0.19 & $-$0.33  \\
          &         &         &         &         &          \\
$[Mn/Fe]$ & $-$1.09 & $-$0.40 & $-$0.06 & $-$0.17 & $-$0.23  \\
          &         &         &         &         &          \\
$[Sr/Fe]$ & $-$0.48 & 0.92    & 0.06    & $-$0.32 & $-$1.52  \\
          &         &         &         &         &          \\
$[Ba/Fe]$ & $-$0.93 & 0.89    & $-$0.01 & $-$0.89 & $<-$1.37 \\
 & & & & & \\ \hline
\end{tabular}
\end{flushleft}
\end{table}

\subsection{Comparison with previous studies}

CS 22968--14 has been analyzed also by Norris et al. (1993),
who found an iron metallicity of [Fe/H] = $-$3.77$\pm$0.26.
The difference between the two metallicities, although
within the errors, can be entirely
explained by the difference in the atmospheric parameters
adopted. In particular we used a T$_{\rm eff}$ of 150\,K higher
and a microturbulence of 0.6\,km\,s$^{-1}$ lower than Norris
et al. (1993). Using Table 6 of their paper, which
investigates the changes in the abundances with the change of
the input stellar parameters, we can see that the differences
between the two atmospheric parameters translate into a
difference in iron of 0.34 dex when summed quadratically, and
we measure an abundance which is 0.32 higher. The same
exercise with all the other elements in common eliminate or
reduce the differences between the two analyses within 0.2 dex
or less. Slightly larger discrepancies are found only for Mn
and are produced from differences in the equivalent widths. The
lines used in the abundance determination of Mn are saturated
and a small difference in the equivalent width translate into
considerable abundance variations. Sodium has been measured in
this star for the first time and the lines are shown in Fig.\,3. \\
For CD $-$38$^{\circ}$245 we found an excellent agreement with the
previous determinations. As it is possible to see from Fig.\,4,6,7
and 8, where the various abundances are connected, in several
cases the abundances we derived here are close to the average
of the previous determinations. From the comparison of the four
abundance determinations available it appears that the low values
for Ti and Al derived by Gratton and Sneden (1988) and by Peterson
et al. (1990) respectively deviate significantly from the other
evaluations.

\begin{figure}
\picplace{18.5cm}
\caption{$\alpha$-elements abundances: (a) magnesium;
(b) silicon; (c) calcium; (d) titanium. {\it Filled circles}
represent our data. The data taken from literature are from:
Carney and Peterson (1981, {\it squares}), Bessel and Norris
(1984, {\it triangles}), Gratton and Sneden (1987, {\it stars};
1988, {\it spiked squares}), Magain (1987a, {\it open circles};
1989, {\it plus signes}), Gilroy et al. (1988, {\it crosses}),
Molaro and Castelli and Molaro and Bonifacio (1990, {\it pentagons}),
Peterson et al. (1990, {\it squares}), Ryan et al. (1991, {\it open
circles}), Norris et al. (1993, {\it diamonds})}
\end{figure}
\subsection{Even light elements (Mg, Si, Ca, and Ti)}

The even-numbered light elements synthesized through $\alpha$
capture processes are generally overabundant in metal-poor
stars resulting from nucleosynthesis of Type\,II supernovae
of stellar masses in the range 10 $\leq$ M/M$_{\sun}$ $\leq$ 30
(Wheeler et al., 1989). Our data (see Fig.\,4) confirm the general
overabundance of the $\alpha$-elements. An increase is possible,
but hidden within the errors, for Mg which is overabundant by 0.73
in CS 22891--209 and together with the values found by Molaro and
Bonifacio (1990) in CS 22881--39 and CS 22885--96 shows an
overabundance which is somewhat larger than the average value of
0.5\,dex found for more metallic stars.
Silicon is particularly constant at $\sim$0.5\,dex over the entire
range of metallicity considered. Calcium abundances look rather
scattered but the dispersion is likely due to the use of the Ca\,{\sc i}
4226\AA~ resonance line. Ryan et al. (1991) found that this strong
line gives systematically lower abundances of about 0.25\,dex. In
our stars the calcium abundances have lower values and relay almost
entirely from the Ca\,{\sc i} 4226\AA~, with the exception of
CS 22891--209 where [Ca/Fe] is 0.43 but averaged among four lines.
Titanium remains overabundant at the lowest metallicities,
but approximately constant without following the increase observed
between [Fe/H] = $-$1.0 and $-$2.5.

\begin{figure}
\picplace{10.5cm}
\caption{Spectral region around the Al resonance lines for CS 22948-66}
\end{figure}
\subsection{Odd light elements (Na and Al)}

The abundance of aluminum has been performed through the resonance
doublets of Al\,{\sc i} at $\lambda\lambda$ 3944\AA, 3961\AA~ for all
the four stars, while the Na\,{\sc i} resonance lines at $\lambda\lambda$
5889\AA, 5895\AA~ are available only for CS 22968--14 and CD
$-$38$^{\circ}$245. Aluminum lines are known
to exibite a puzzling behaviour often showing strengths opposite
to the theoretical ones. The line at 3944\AA~ is blended with CH
and so, following the usual approach, we have derived the abundance
from the 3961\AA~ line.  Figure\,5 reproduces the spectral regions
around these two lines for CS 22948--66.

Carbon burning theory predicts Na and Al underabundant in
comparison with the $\alpha$-elements (the {\it odd-even\/} effect).
The synthesis of odd atomic number elements is sensitive to the
neutron excess and neutrons are essentially supplied by metals.
In metal-poor gas there is a lower neutron excess and consequently
lower abundances of the odd elements (Woosley and Weaver, 1982;
Woosley and Weaver, 1986).
Figure\,6b shows a clear underabundance of aluminum with respect to
iron down to [Fe/H]$\approx$$-$3.0, while at lower metallicities the
abundance rises again, and in CS 22891--209 it becomes solar
with [Al/Fe] of $-$0.03\,dex. This value has to be taken with some
caution since errors may be large. Decreasing the equivalent width
by 20 m\AA~, i.e. of 1$\sigma$, gives an abundance 0.4\,dex lower.
Sodium is rather controversial with a considerable lack of data
at the lowest metallicities but showing little evidence for a
falloff in the abundances at low metallicities (see Fig.\,6a).
The sodium lines for CD $-$38$^{\circ}$245 and CS 22968--14 are
reproduced in Fig.\,3, together with the synthetic spectrum.

The presence of the {\it odd-even\/} effect  is more properly seen
when the abundances of the odd-elements sodium and aluminum are
compared to that of magnesium, because they are produced in the
same burning phase. For instance, while [Al/Fe] is solar in CS
22891--209, [Al/Mg] is clearly underabundant since this star shows
a particularly high overabundance of magnesium. When compared
to magnesium, the rise seen in [Al/Fe] for [Fe/H] $\leq$ $-$3.0
becomes a plateau around $-$1.0 for [Mg/H] $\leq$ $-$2.0 (see
Figs.\,6c-d). The trend is even more evident if we do not consider
the relatively low abundance for aluminum in CD $-$38$^{\circ}$245
found by Peterson et al. (1990), which is found considerably off
from the other three determinations. The present result strengthen
early indications found by Bessel and Norris (1984), Molaro and
Castelli (1990), Ryan et al. (1991) and Norris et al. (1993).
Woosley (see Bessel and Norris, 1984 -- footnote 2), to explain the
anomalous high Al abundance observed first in CD $-$38$^{\circ}$245
by Bessel and Norris (1984), pointed out that odd-element production
follows from a direct neutron production during the CNO cycle, which
becomes the dominant source of neutrons at very low metallicities.
The observed plateau thus strongly suggests the presence of a primary
behaviour for aluminum and a suppression of the odd-even effect for
[Fe/H] $\leq$ $-$3.0 or [Mg/Fe] $\leq$ $-$2.0.
\begin{figure}
\picplace{20cm}
\caption{Odd-numbered light elements Al and Na, compared with iron (a-b)
and magnesium (c-d). For the symbols see the list given in the caption of
Fig.\,4}
\end{figure}

\begin{figure}
\picplace{14cm}
\caption{Iron-peak elements abundances: (a) scandium; (b) manganese; (c)
cromium. For the symbols see the list given in Fig.\,4}
\end{figure}
\subsection{Iron-peak elements (Sc, Mn and Cr)}

In Fig.\,7 we report the trends for scandium, manganese and chromium.
The abundances of these iron-peak elements follow closely the iron
abundance with relative abundances essentially solar over the
entire metallicity range explored, and this is also for our four
objects. The odd-numbered scandium shows a slight overabundance with
weak evidence of odd-even effect compared to the neighbouring elements
Ti and, perhaps, Ca. Manganese and iron are an odd-even pair and Norris
et al. (1993) found an odd-even effect of the order of 0.4\,dex, which
we can confirm only for CS 22891--209. Our manganese abundances with
[Fe/H]$\sim$ $-$3.0 are aligned with those of stars with similar
metallicities, in particular with the dwarfs of Ryan et al. (1991).
It is possible that a real scatter in the manganese abundance might occur
at [Fe/H] $\approx$ $-$4.0. Here we have CS 22885--96 with [Mn/Fe]
$\sim$ $-$0.34 (Norris et al., 1993), but CD$-$38$^{\circ}$245 which we
confirm here with [Mn/Fe]$\sim$ $-$1.09 and CS 22876--32 with [Mn/Fe]
$\sim$0.46 (Molaro and Castelli, 1990). Chromium is characterized by a
nearly constant underabundance along the entire metallicity range of
the halo, which becomes more noticeable for CD $-$38$^{\circ}$245.

\begin{figure}
\picplace{9.5cm}
\caption{Heavy elements abundances: (a) strontium; (b) barium. For the
symbols see the list given in Fig.\,4}
\end{figure}
\subsection{Heavy elements (Sr and Ba)}

Heavy elements are mainly synthesized through neutron capture
reactions by iron-peak nuclei which are exposed to {\it slow\/}
or {\it rapid\/} neutron fluxes.
The {\it r\/}-process is essentially occurring in the interiors of
Type\,II supernovae, while the {\it s\/}-process is mainly
occurring both in massive and low-mass stars during the He-burning
phase or thermal-pulses (Baraffe and Takahashi, 1993).
Stars which are sites for the {\it s\/}-process do not produce
seed nuclei (typically Fe) which must have existed at stellar
birth. The {\it s\/}-process is therefore secondary and the
expected relation between {\it s\/}-elements and iron is linear,
[{\it s\/}/Fe] $\propto$ [Fe/H] (Baraffe and Takahashi, 1993).
The {\it r\/}-process operates on seed nuclei freshly made and
therefore is a primary process with the resulting abundance ratios
unrelated to the stellar metallicity (Mathews and Cowan, 1990).
Truran (1981), from the observed behaviour of the abundance of Ba
and Y for decreasing metallicities, suggested that the heavy elements
observed in the Pop II stars are {\it r\/}-process products even
for those elements dominated by the {\it s\/}-process in the solar system.
Moreover, a possible change of slope in [Ba/Fe] versus [Fe/H] around
[Fe/H]$\sim$ $-$2.0, which can be seen in Fig.\,8 from the literature
data, could imply a contribution by the {\it s\/}-process in low- and
intermediate-mass stars (Baraffe and Takahashi, 1993).
Also europium shows a decline for [Fe/H] $\leq$ $-$2.0 though
essentially solar at greater metallicities (Gilroy et al., 1988).
However, Eu is 90\% {\it r\/}-element in the solar system and the decline
has been interpreted as a dependence of its production on the mass
of the star (Mathews et al., 1992). \\
Our analysis has been restricted to strontium and barium through
the strong lines of Sr\,{\sc ii} at $\lambda\lambda$ 4077\AA, 4215\AA,
and Ba\,{\sc ii} at $\lambda\lambda$ 4554\AA, 4934\AA, which are the
only transitions detectable in our spectra. Both the elements are
detected in all the stars but CS 22968--14, where Sr is only
marginally detected with [Sr/Fe] = $-$1.52 and no barium lines are
visible providing [Ba/Fe] $\leq$ $-$1.37 (see Figs.\,9-10).\\
{}From Fig.\,8 it is possible to see that the abundance we derive
in our four stars does not follow strictly the trends shown
from the data of the literature. In particular, the data for Ba are
not consistent with a monotonic decline of the abundance of this
element with the decrease of metallicity. In CS 22897--8 [Ba/Fe] is
about solar and in CS 22891--209 is overabundant by 1.0 dex. Few
other stars with [Ba/Fe] close to solar or in excess have been found
by Ryan et al. (1991) and Norris et al. (1993) around [Fe/H] $\approx$
$-$3.0 as Fig.\,8b shows. For Sr our data does not support a complete
falloff of its abundance at least down to [Fe/H] $\sim$ $-$3.4.\\
The most striking feature emerging from our measurements is the
presence of a large scatter among the abundance of Sr and Ba. The
differences in [Sr/H] and [Ba/H] between CS 22871--209 and CS 22968--14
are of $\sim$2.6\,dex (see Table 3) and exceed largely the associated
errors which are of $\pm$0.6\,dex at most for each star (cfr. Table 6).
A particularly large dispersion has already been noted by Ryan et al.
(1991) and Norris et al. (1993) on stars of [Fe/H]$\simeq$ $-$3.0, and
also, but with a smaller amplitude and on more metallic stars, by Gilroy
et al. (1988). The dispersion in our data points is about twice that
found by Gilroy et al. (1988), providing some evidence that the scatter
becomes larger as the metallicity decreases.
Stellar processes are unlikely to be responsible for the observed dispersion,
which therefore must reflect inhomogeneities in the protostellar nebula.
Thus the dispersion observed in the heavy elements Sr and Ba is rather
peculiar among the elements and it may be an important signature of their
production mechanisms. The two explanations putforward by Mathews et al.
(1992) and Woosley and Hoffman (1992) to explain the falloff of Eu may
lead as well to chemical inhomogeneities.
Mathews et al. (1992) argued that stars with different masses may have
different heavy elements outputs with the more massive supernovae
producing less {\it r\/}-elements.
Woosley and Hoffman (1992) suggested instead that the Eu behaviour is
possibly the result of mass-dependent iron production but with a
mass-independent {\it r\/}-production.
Both these scenarios result with a mass-dependent [{\it r\/}/Fe] and
if the mixing was not very efficient in comparison with the very short
timescale of the first elemental build-up they may provide also a
plausible mechanism to explain the observed dispersion.
\begin{figure}
\picplace{9.5cm}
\caption{Spectral region around Sr{\sc ii} 4077\AA}
\end{figure}
\begin{figure}
\picplace{14cm}
\caption{Spectral region around Ba{\sc ii} 4554\AA}
\end{figure}

\newpage
\begin{table*}
\caption{Computed U$-$B and B$-$V colours for [M/H]=$-$3.0}
\picplace{11cm}
\end{table*}
\begin{table*}
\caption{Computed U$-$B and B$-$V colours for [M/H]=$-$3.5}
\picplace{11cm}
\end{table*}
\section {Appendix A}

\subsection{Computed colour indices U$-$B and B$-$V}

By using the ATLAS9 code (Kurucz, 1993a), the opacity distribution
functions M35ABIG2, M35ALIT2, M30ABIG2, M30ALIT2 (Kurucz, 1993a)
and the UBVBUSER code (Kurucz, 1993b) we computed a small grid of
models, fluxes, and Johnson UBV colours.
The abundances are solar scaled [M/H] = $-$3.5 and [M/H] = $-$3.0
for all the elements except for the $\alpha$-elements O, Ne, Mg,
Si, S, Ar, Ca and Ti, for which an abundance enhanced by 0.4\,dex
was assumed.
Solar abundances used for reference are those of Anders and Grevesse
(1989), except for iron for which the value
$\log$ N$_{\rm Fe}$/N$_{\rm tot}$ = $-$4.53 (Hannaford et al., 1992) was
assumed.
The microturbulent velocity in the opacity tables is equal to
2\,km\,s$^{-1}$. Convection was treated with the mixing-length theory,
taking the mixing-length to the pressure scale-height ratio l/h equal
to 1.25. No overshooting was considered.
All models converged with flux errors and flux derivative errors lower
than 1\% for nearly all the 72 layers considered, which range from
$\log\tau_{\rm Ross}$ = $-$6.875 to 2.000.
Only in a few cases and for a few layers the errors were of the order
of 1\%.
Tables 8 and 9 list the U$-$B and B$-$V colours computed for models
having T$_{\rm eff}$ ranging from 4500 to 5750\,K and $\log g$ ranging
from 1.0 to 2.0, with [M/H] = $-$3.0 and $-$3.5 respectively.
The computed indices are transformed to the standard system by using
as zero point the indices of Vega and the Vega model of Castelli and
Kurucz (1994).

\end{document}